# Spectral Density on the Lattice

D. Makovoz

Department of Physics, 174 W. 18th Av., Ohio State University, Columbus, OH 43210, USA

G. A. Miller

Department ofPhysics, FM-15, University of Washington, Seattle, WA 98195, USA

November 7, 1995

**Abstract**

Spectral density in the pseudoscalar and vector channels is extracted from the SU(2) lattice quenched data. It is shown to consist of three sharp poles within the energy range accessible on the lattice.

# 1 Introduction

Lattice QCD has provided many results for ground state hadronic properties[1, 2]. In a number of papers an attempt has been made to analyze the first excited state as well [3, 4]. The general strategy is to create a source with a good overlap with a particular state and then look for a plateau in the Euclidean time to extract the properties of the selected state. Thus the information from the shorter time is usually discarded since its analysis requires some knowledge of the spectral density function. However, it is a well known fact that the short time data suffers the least from the statistical error[5].

The ratio of the statistical error $\sigma(t)$ for a hadron-hadron correlation function $G(t)$ to the signal is expected [6] to increase with $t$ in any non-pion channel:

$$\frac{G(t)}{\sigma(t)} \sim \frac{\exp -Mt}{\exp -n_q m_\pi t/2}, \qquad (1)$$

where $m_\pi$ is the mass of the pion, $n_q = 2$ (3) for mesons (baryons), $M$ is the mass of the lowest-lying state in the channel under consideration. Moreover, a recent study [7] showed that in the pion channel the "effective mass as a function of time should not exhibit long plateaux whatever high statistics simulation are made". This emphasizes the importance of understanding the phenomena occuring at short times.

Two recent papers [8, 9] examine the short time (distance) region using a form of the spectral density based on a theory of free massless quarks. It would be very advantageous to provide more detailed models for the spectral density to represent lattice data for short times. The rewards are twofold. If successful, one would be able to extract spectral density functions from lattice calculations. The spectral density is closely related to experimental scattering data [10]. The results of lattice calculations could be compared with available data and complement experiment in the channels where data are unavailable. Even if a conclusive decision about the form of the spectral density can not be made, a good parametrization will allow for better determination of the ground state properties using the short time information.

In our previous paper [11] we addressed the question of lattice QCD theories of the spectral density. The conclusion reached in [11] was, that for SU(2) color, the spectral density in the accessible range of energy is well approximated by the set of three poles. This result is based on an analysis of data for the two-point zero-momentum correlation functions in the pion channel. The present paper is devoted entirely to the issue of spectral density and goes beyond that work in several ways. We improve our statistics for the heaviest quark mass. We perform a careful analysis of the results for both the pseudoscalar and vector channels. Two classes of correlation functions, the spacial correlation function $S_2(x)$[9] and the zero momentum two-point correlation function $G_2(t)$[8] are examined. Questions related to the stability of the results of the multipole fitting and reliability of our fitting technique are examined in great detail.

In Section 2 we introduce the three parametrization of the spectral density functions used in this paper. In Section 3 the details of the lattice calculations are given. In Section 4 results for the zero-momentum correlation functions are presented and discussed. We measure $G_2(t)$ in vector and pseudoscalar channels for three $\kappa$'s. Then we perform a correlated $\chi^2$ fit of the measured functions with the theoretical form given by the three model spectral density functions. A detailed comparison of the three parametrization functions of the spectral density in the pseudoscalar and vector channel is given. We show that the three pole spectral density describes our results the best, and argue further that not only is the three pole model the best parametrization of our spectral density, but that within the present resolution and energy range the spectral density in the two investigated channels is actually discrete and consists of the three poles. In Section 5 we present analogous calculations for the spatial correlation function $S_2(x)$, which is measured in both channels for two $\kappa$'s and fit with the first two forms of the spectral density function. The pole plus continuum form completely fails to fit $S_2(x)$, the many pole form fits it poorly. The suggested reason is the large systematic error caused by the lattice anisotropy.

The problem with using lattice calculations to examine the spectral density is that the contribution of the high energy part of the spectrum is not readily accessible. Due to Euclidean time decay the contributions from states at high energy very quickly go below the statistical error. In Section 6 possible ways to improve sensitivity to the high energy part of the spectral density are discussed. We summarize our results in Section 7.

## 2  Formalism

The most straightforward way to obtain the information needed to extrapolate the spectral density function is to measure the two-point correlation functions $S_2^h(x)$ defined as the vacuum expectation value of the time-ordered product of the interpolating fields

$\phi_h(x)$

$$S_2^h(x) = <0|\hat{T}(\phi_h(x)\phi_h(0))|0>, \quad (2)$$

or its three-dimensional Fourier-transform definite momentum correlation function $G_2^h(t,\vec{p})$

$$G_2^h(t,\vec{p}) = \sum_{\vec{x}} e^{-i\vec{p}\cdot\vec{x}} <0|\phi_h(\vec{x},t)\phi_h(0,0)|0>. \quad (3)$$

Here the four-vector $x$ is defined as $(\vec{x},t)$.

The following interpolating fields are used in the pseudoscalar and vector channels:

$$\phi_\pi(x) = \bar{d}(x)\gamma_5 u(x) \quad (4)$$
$$\phi_\rho(x) = \bar{d}(x)\gamma_1 u(x).$$

For a local field theory the two-point correlation function in momentum space can be uniquely characterized up to a polynomial by the spectral density $f(s)$

$$\int d^4x e^{iqx} <\phi_h \phi_h> = \int ds \frac{f(s)}{s-q^2-i\varepsilon}. \quad (5)$$

In the case of free massless quarks the spectral density f(s) can be calculated easily. In a mesonic channel $f(s) \sim s$, the form related to the experimental cross section of $e^+e^- \to$ hadron production. We do not in general know how to calculate the spectral density function for interacting quarks. We can use experimental information and general properties of local field theories to derive its qualitative behavior. We do know that there is usually a sharp resonance corresponding to the ground state in the channel under consideration. Looking at the long time tail of the zero-momentum correlation function is equivalent to using only this part of the spectral density parameterized as a pole term.

There have been a few papers where an attempt has been made to use a specific parametrization of the spectral density to examine the short distance region([8, 9]). They used f(s) consisting of a pole plus a continuum inspired by the free massless quark spectral density:

$$f(s) = \lambda^2 \delta(s-E_1^2) + c_{cont} s \theta(s-s_0). \quad (6)$$

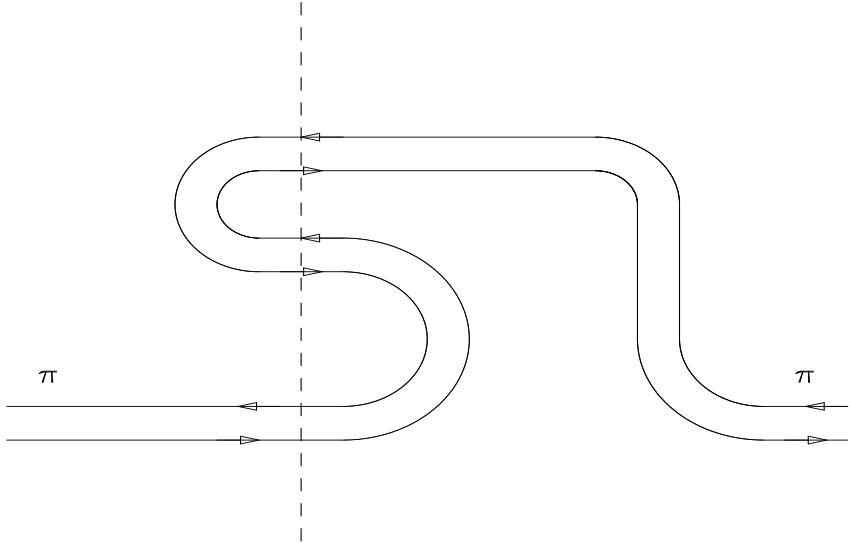

Figure 1: Backward-going fermion lines create a state with the quantum numbers of three pions.

The goal of this paper is to investigate the form of the spectral density function on the lattice in the quenched approximation. We use both the spatial correlation function $S_2(x)$ [9] and the 0-momentum two-point correlation function $G_2(t)$ [8]. We also use two more parametrization for the spectral density.

The first one is the sum of several sharp poles:

$$f(s) = \sum_n \lambda_n^2 \delta(s - E_n^2), \qquad (7)$$

and the second one consists of a sharp pole plus several "resonances" characterized by the finite widths $1/a_n$

$$f(s) = \lambda_1^2 \delta(s - E_1^2) + \sum_{n=2,\ldots} \lambda_n^2 a_n e^{-(2\sqrt{s} - E_n)^2 E_n a_n^2}. \qquad (8)$$

The rational behind the sharp pole plus several "resonances" form is that, due to multiparticle excitations, stable particles turn into broad and often overlapping resonances. The reason to expect a discrete spectral density is the quenched approximation. Let us look at the large N limit([12, 13]) of QCD, which is even stronger assumption than that of quenching. Not only are there no internal quark loops, but there are only planar diagrams. The coupling of a meson state to a multi-meson state or to a meson +glueball state $\sim 1/N^\gamma$, where $\gamma \geq 1$. The meson spectrum is discrete. There are no decays of a meson into many mesons or into a meson plus a glueball. To the extent that the quenched approximation is similar to the large N limit we expect to get the same conclusion about the spectral density. But let us examine the diagram in Fig 1, which indicates that quenched QCD might have multi-pion states. The intermediate state contains a $3q\bar{q}$ structure which suggests that there is a state of three pions. At the start, we do not know whether there is any spectral strength corresponding to a $3\pi$ state, or this diagram represents a multi-quark component of a one $\pi$ wave function. Whichever is the case, we shall use the discrete form of the spectral density here and shall do it to successfully model the lattice data.

## 3 Lattice Details

The lattice calculations are performed on the UW Nuclear Theory DECstation 3000-600 AXP (ALPHA) with maximum speed of 140 Mflops.

The calculations are performed for $SU(2)$ gauge theory with the coupling constant $\beta = 2.5$. $SU(2)$ is chosen to increase the efficiency of the calculation and improve statistics. The size of the lattice is $12^3 \times 24$. 360 quenched gauge field configurations are generated using Metropolis method with overrelaxation [21]. The first configuration is selected after 2000 thermalization sweeps and all the consecutive ones after 1000 sweeps. The calculation of each configuration took approximately 2.5 hours of CPU time.

The quenched approximation is used. The Wilson propagators are calculated for three values of the hopping parameter $\kappa = 0.146, 0.148, 0.149$. Periodic boundary condition in the spatial directions and Dirichlet boundary conditions in the time direction are used. The inversion time per configuration ranged from approximately 2 hours of CPU time for $\kappa = 0.146$ to 3 hours of CPU time for $\kappa = 0.149$.

We calculate two sets of propagators with different source locations. The first set has the source at time slice $t = 5$. These propagators are used to form the definite momentum correlation functions $G_2(t)$. 120 propagators are calculated for the hopping parameters $\kappa = 0.148, 0.149$. For the hopping parameter $\kappa = 0.146$ we calculate 360 propagators. The second set has the source at time slice $t = 12$ in the middle of the lattice. These propagators are used to form the spatial correlation functions $S_2(x)$. The source is in the middle to minimize the influence of the boundaries in the time direction. 120 propagators are calculated for the hopping parameters $\kappa = 0.146, 0.149$.

## 4 Zero-Momentum Correlation Functions

The zero-momentum correlation function $G_2(t)$ has the following spectral representation:

$$G_2(t) = \int_0^\infty \frac{ds}{2\sqrt{s}} f(s) e^{-\sqrt{s}t}, \qquad (9)$$

which follows from the definition of the two-point correlation function (3) and the dispersion relation (5).

We use the three forms of the spectral density function (6), (7), and (8) to calculate the fit functions to fit the measured two-point correlation function. Thus we get the many pole fit function

$$G_p(t) = \sum_n c_n^2 e^{-E_n t}, \qquad (10)$$

the pole plus continuum fit function

$$G_c(t) = c_1^2 e^{-E_1 t} + c_{cont} e^{-\sqrt{s_0}t} \left( \frac{1}{t^3} + \frac{2\sqrt{s_0}}{t^2} + \frac{2s_0}{t} \right), \qquad (11)$$

and the pole plus several resonances fit function

$$G_g(t) = c_1^2 e^{-E_1 t} + \frac{\sqrt{\pi}}{2} \sum_n c_n^2 (1 - erf(\frac{t}{2a_n} - a_n E_n)) e^{-E_n t + \frac{t^2}{4a_n^2}}, \qquad (12)$$

where $erf(t)$ is the error function, and $c_n \equiv \lambda_n / \sqrt{2E_n}$.

What follows is a recalculation of the work of [11] based on data with better statistics. The number of configurations has grown from 120 to 360. To obtain the parameters of the fit functions we perform the correlated $\chi^2$ fit using the full covariance matrix [14]. We use $G_2(t)$ for $\pi$ and $\kappa = 0.146$ as an example, since for these hopping parameter we have three times as much configurations as for the other two. The results for other values of $\kappa$ and for $\rho$ are very similar. The fitting is performed over a time range extending from $t_{first}$ to $t_{last}$. We choose $t_{last} = 20$ to exclude the boundary effects and $t_{first}$ is varied from 6 to 14. We also use quantity $\Delta t = t_{first} - t_{source}$ to describe the distance from the source, located at $t_{source} = 5$, to the first time slice included in the fit.

There have been studies [16, 15] to show that for a small data sample the use of the full covariance matrix is unreliable. Even the number of configurations $N_{conf} = 120$ we use for $\kappa = 0.148$ and $0.149$ satisfies the criterion proposed in [16] that

$$N_{conf} > 10(D+1), \tag{13}$$

where $D$ is the number of the degrees of freedom, equal to the number of the time slices used in the fitting minus the number of fit parameters.

First we fit the data with the many pole form $G_p(t)$. The result depends on the $t_{first}$ used in the fit. For sufficiently large $t_{first}$ only the contribution of the ground state is significant. As we get closer to the source the calculations become sensitive to higher energies in the spectral density. A new state has to be included in the fit. This is done based on $\chi^2/dof$. The results for the $E_n$'s for $\pi$ for the hopping parameter $\kappa = 0.146$ are shown in Fig. 2. For $t_{first} \geq 13$ one state is enough, for $t_{first} = 9 \div 11$ two states had to be included, and for $t_{first} = 6 \div 8$ the $G_2(t)$ is best fit with three states.

Of course our sensitivity to a particular state is determined by the quality of the data, i.e. by the statistical error which will always be present. In Fig. 3 we show the individual contributions $\lambda_n^2 e^{-E_n t}$ of the states to the two-point correlation function $G_2(t)$ compared to the standard deviation of $G_2(t)$. Two time slices $t_{first} = 13$ and $t_{first} = 9$ are important, since at those times the second and third states, go above the statistical noise. Those are precisely the time slices after which the contribution of these states is included in the fit. Here we want to contrast this result with the one quoted in [11]. In that paper with 120 configurations the statistical error of the data was greater. It is shown in Fig 3. With the better statistics we are able now to determine the second state for $t_{first} = 12$, whereas in [11] the farthest was $t_{first} = 11$. Unfortunately, to increase the $t_{first}$ for the third state from 8 to 9 we need to decrease the statistical error by the factor of $\sim 5$. This translates into $360 \cdot 5^2 = 9000$ configurations, which makes such an endeavor not feasible. In section 6 we discuss an alternative way to improve the sensitivity to the higher excited states.

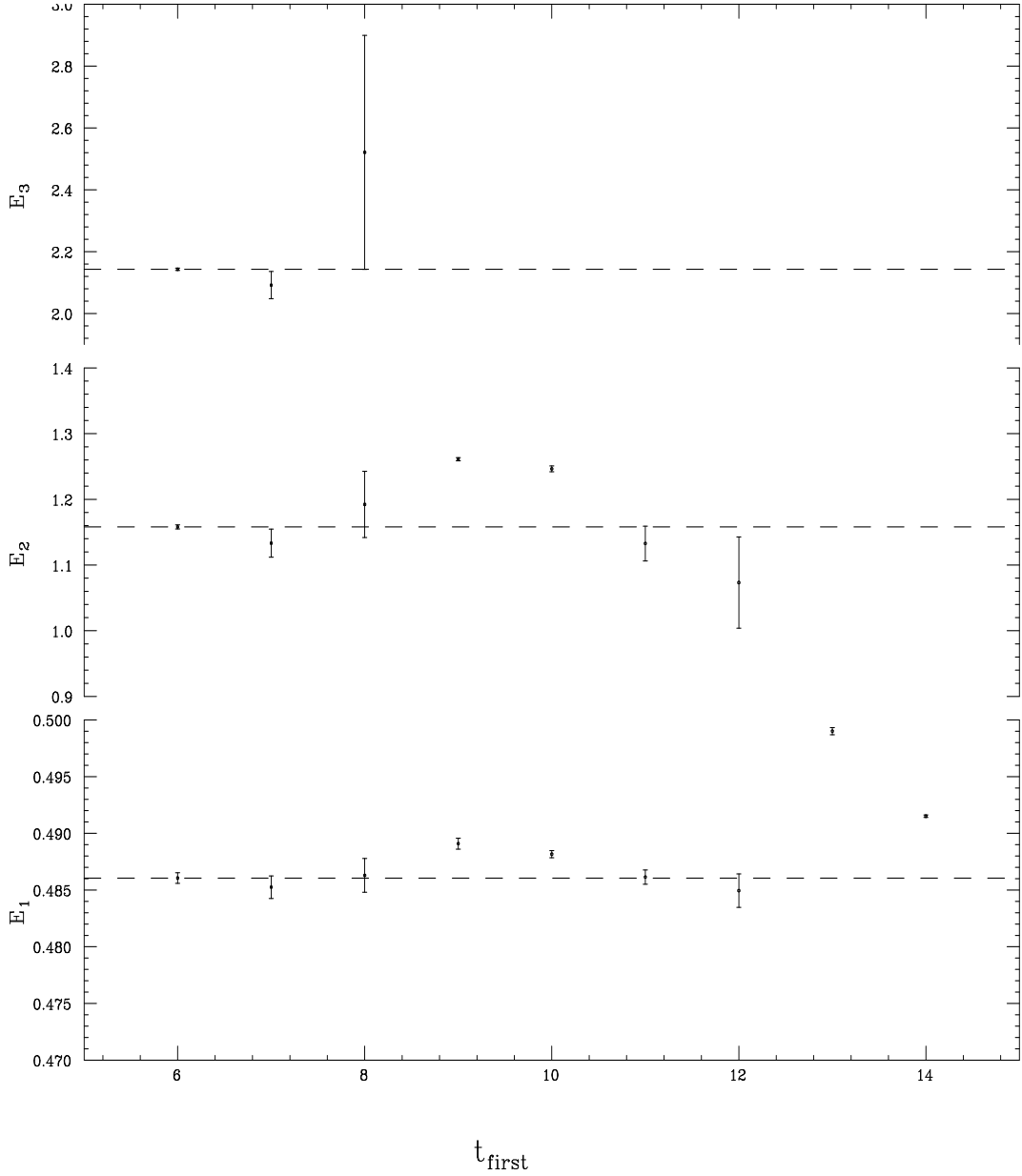

Figure 2: The energies of three states as a function of the first time slice $t_{first}$ used in the fitting. $t_{last} = 20$ for all the fits. The dashed line shows the result for the longest time range. Hopping parameter $\kappa = 0.146$.

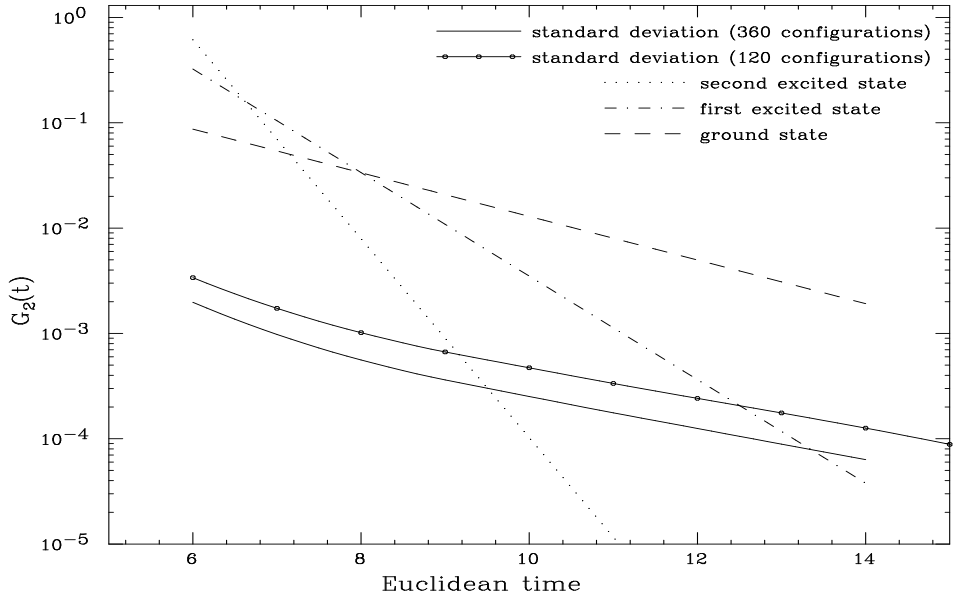

Figure 3: Individual contributions of the three states to the two-point correlation function to be compared to the statistical error of the two-point correlation function $G_2(t)$. Hopping parameter $\kappa = 0.146$.

A state should be included if its contribution is greater than the statistical error for at least two time slices; it should not be included otherwise. This is a modification of the rule stated in our first paper [11]. The best results of the fitting are obtained for the longest time span of the highest energy state used in the fitting. The values of the $\chi^2/dof$ show that the fit for the longest time range is the best( Table 1). In Table 2 we compare the energy and excitation strength for the three states in the $\pi$ channel, $\kappa = 0.146$ for 120 and 360 configurations. [1]

| | $t_{first}$ | 6 | 7 | 8 | 9 | 10 | 11 | 12 |
|---|---|---|---|---|---|---|---|---|
| many pole fit | $\chi^2/dof$ | 0.53 | 0.59 | 0.63 | 0.54 | 0.59 | 0.60 | 0.71 |
| | confidence level | 0.86 | 0.79 | 0.73 | 0.64 | 0.77 | 0.73 | 0.61 |
| pole plus continuum fit | $\chi^2/dof$ | 123 | 0.55 | 0.61 | 0.58 | 0.65 | 0.75 | 0.93 |
| | confidence level | 0.0 | 0.85 | 0.79 | 0.79 | 0.72 | 0.61 | 0.46 |

Table 1: $\chi^2$ per degree of freedom and confidence levels for the many pole model and pole plus continuum model fits; $\kappa = 0.146$.

Result should be consistent if one uses different time ranges. Variations in the parameters with the first time slice could be regarded as systematic errors. But, as we see, fits for some time ranges have more credibility than for others. Consider the contribution of a single state. At least three time slices are necessary to perform a genuine fit of this contribution, since each pole is determined by two parameters. Here is another improvement that we get from the 360 configurations compared to 120. With 360 configurations we can have a genuine fit with three states for two values of $t_{first} = 6, 7$ (as opposed to $t_{first} = 6$ with 120 configurations) and genuine fit with two states for three values of $t_{first} = 9, 10, 11$ (as opposed to $t_{first} = 9, 10$ with 120 configurations).

As a form of a continuum spectral density we use the pole plus two resonances form (8). This form has two remarkable features: (1) a great variety of possible forms of the spectral density can be generated (Fig. 4) by varying the parameters of this model; (2) this model has the three pole form as one of its limits. Three delta functions belong to the same family as other curves shown in Fig. 4 and a continuous transition can be made by taking the limit $a_n \to \infty$. We fit $G_2(t)$ with ( 8) and discover that the sharp poles are preferred over all other possible shapes. It is interesting to note that good statistics is essential for this result. When we use fewer than $\sim 100$ configurations the fit with the pole plus two gaussians does not degenerate into three poles. Only when the number of configurations exceeds 100 do we obtain three poles in this fit. This result does not change when the number of configurations goes up to our maximum of 360.

---
[1] The errors quoted in[11] are slightly higher than those shown here due to a bug in the error calculating code.

| Parameter | 120 configurations | 360 configurations |
|---|---|---|
| $E_1 a$ | 0.477(1) | 0.4861(5) |
| $E_2 a$ | 1.13(1) | 1.158(3) |
| $E_3 a$ | 2.18(1) | 2.143(4) |
| $c_1 a^{3/2}$ | 0.374(2) | 0.387(1) |
| $c_2 a^{3/2}$ | 1.00(1) | 1.006(3) |
| $c_3 a^{3/2}$ | 2.32(1) | 2.29(1) |

Table 2: Energy $E_n$ and excitation strength $c_n$ of the three pole fit of the two-point correlation function $G_2(t)$; $\kappa = 0.146$.

Another advantage of having better statistics is that the number of time slices where the contributions of the excited states is above the statistical error is now 4 for the second excited state and another 4 for the first excited state ( see Fig. 4). There are three parameters per state in the Gaussian model, and this allows an accurate fit to the data.

We see that one can fit the lattice data with the three pole model. How about the pole plus continuum model? We do not need the strategy we used for the many pole fit since the same set of parameters describes the continuum from $s_0$ to infinity. Nevertheless, we still require the results for different time ranges to be consistent. It comes as a suprise that the pole plus continuum model fails miserably when we use the longest time range with the time slice next to the source is included into the fit. We do not obtain the correct values of the ground state parameters and $\chi^2/dof = 137$.

So far we have only repeated the results of paper [11] obtained for the data with better statistics. We shall now give a more careful analysis of the fitting procedure and results.

In [8] it was suggested that for small distances we need to explicitly include the ultraviolet lattice cut-off. This cut-off can be modeled by introducing the upper limit $s_{cut}$ in the integral (9) over in energy

$$G_2(t) = \int_{s_0}^{s_{cut}} \frac{ds}{2\sqrt{s}} f(s) e^{-\sqrt{s}t}. \tag{14}$$

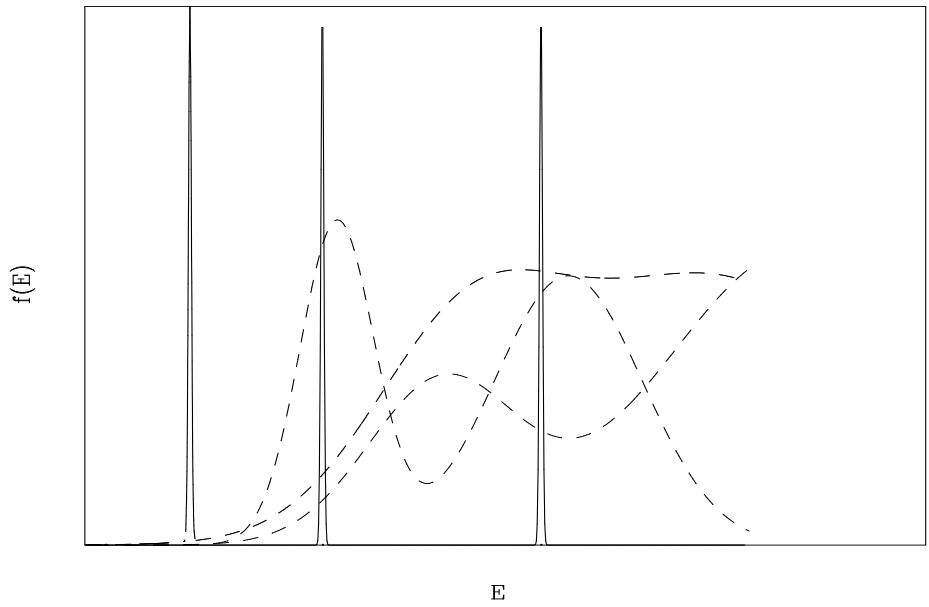

Figure 4: The spectral density for the pole plus two Gaussians model. Dashed lines show possible shapes obtained for several sets of the parameters. The solid line symbolically represents the three delta-functions.

With this cut-off the $G_c(t)$ is modified to

$$G_c(t) = c_1^2 e^{-E_1 t} + c_{cont} e^{-\sqrt{s_0} t} \left(\frac{1}{t^3} + \frac{2\sqrt{s_0}}{t^2} + \frac{2s_0}{t}\right) - \qquad (15)$$
$$c_{cont} e^{-\sqrt{s_{cut}} t} \left(\frac{1}{t^3} + \frac{\sqrt{2s_{cut}}}{t^2} + \frac{2s_{cut}}{t}\right).$$

In that paper the author used the operator product expansion approach to model the spectral density to fit the lattice data. Based on this approach the next term in the OPE proportional to the quark mass has to be included for short times $t$. With this modification $G_c(t)$ takes on the following form:

$$G_c(t) = c_1^2 e^{-E_1 t} + c_{cont} e^{-\sqrt{s_0} t} \left[\left(\frac{1}{t^3} + \frac{2\sqrt{s_0}}{t^2} + \frac{2s_0}{t}\right) + \qquad (16)\right.$$
$$c_{mq}\left(\frac{1}{t^2} + \frac{\sqrt{s_0}}{t}\right)\bigg] -$$
$$c_{cont} e^{-\sqrt{s_{cut}} t} \bigg[\left(\frac{1}{t^3} + \frac{2\sqrt{s_{cut}}}{t^2} + \frac{2s_{cut}}{t}\right) +$$
$$c_{mq}\left(\frac{1}{t^2} + \frac{\sqrt{s_{cut}}}{t}\right)\bigg],$$

where $c_{mq} \approx m_q a$, and its exact value to be determined from the lattice calculations, $m_q$ is the quark mass.

The results of fitting $G_2(t)$ with (6), (15), and (16) are given in Table 3. None of the models fit the data with $t_{first} = 6$ ($\Delta t = 1$) included into the fit. Even though model (16) with the quark mass correction has much lower $\chi^2/dof$ the values of the parameters are unphysical and completely inconsistent with the ones obtained for shorter time ranges. Moreover, even for shorter time ranges the quark mass correction model yields unphysical value for $c_{mq}$, which is expected to be $\sim m_q a \sim 0.1$. Inclusion of the energy cut-off seems to improve the quality of the fit without giving any unphysical values for the parameters. Our data is not really sensitive to the spectral density around the upper limit, so we do not ascribe any significance to the exact value of $s_{cut}$ obtained. We use the explicit cut-off only for the fits with $t_{first} =6$ and 7 ($\Delta t = 1$ and 2).

It was pointed out in [3] that high values of $\chi^2/dof$ may not be the result of a poor fit, but rather can be caused by the ill-defined low-lying eigenmodes of the correlation matrix. It was proposed to use several well defined eigenmodes of the covariance matrix instead of the full matrix. In most of our fits we obtained very low values of $\chi^2/dof$. We test the technique proposed in [3] to see if the high values of $\chi^2/dof$ mentioned above are caused by the ill-defined low-lying eigenmodes of the covariance matrix. The lattice data $G_2(t)$ is fit with the pole plus continuum and a cut-off model (15). The covariance matrix is diagonalized and several low-lying eigenmodes are omitted when $\chi^2$ is calculated. The result is that the value of $\chi^2$ remains virtually the same when the number of modes is decreased from its maximum of 16 to the minimum of 5.

| model | $\Delta t$ | $c_1^2$ | $E_1$ | $c_{cont}$ | $c_{mq}$ | $\sqrt{s_0}$ | $\sqrt{s_{cut}}$ | $\chi^2/dof$ |
|---|---|---|---|---|---|---|---|---|
| | 1 | 2.05 | 1.19 | 0.25 | - | 1.70 | - | 187 |
| p.c. | 2 | 0.13 | 0.47 | 1.23 | - | 0.80 | - | 1.52 |
| | 3 | 0.14 | 0.48 | 1.26 | - | 0.86 | - | 0.63 |
| | 1 | 0.47 | 0.73 | 6.1 | - | 1.72 | 1.96 | 122 |
| p.c.c. | 2 | 0.15 | 0.48 | 1.26 | - | 0.88 | 3.76 | 0.55 |
| | 3 | 0.14 | 0.48 | 1.25 | - | 0.87 | 2.72 | 0.61 |
| | 1 | 0.17 | 0.49 | 0.03 | 45 | 1.09 | 4.33 | 1.45 |
| p.c.c.m. | 2 | 0.15 | 0.48 | 0.57 | 1.16 | 1.0 | 8.15 | 0.55 |
| | 3 | 0.15 | 0.48 | 0.51 | 1.43 | 0.98 | 5.29 | 0.57 |

Table 3: Fit of the lattice data with: pole plus continuum (p.c.), pole plus continuum and a cut-off (p.c.c), pole plus continuum with a cut-off and the term proportional to the quark mass (p.c.c.m.). $\Delta t$ is the distance from the source in the time direction.

Results for $E_1$, $\sqrt{s_0}$, and $\sqrt{s_{cut}}$ for various time ranges are shown in Fig. 5. In Fig. 6 we show how the three pole and pole plus continuum models fit the data. The parameters for the three pole model are taken from the fit with $t_{first} = 6$ ($\Delta t = 1$) and for the pole plus continuum model from the fit with $t_{first} = 7$ ($\Delta t = 2$). The graph gives a clear illustration how both models fit the data really well, except for the time range with $t_{first} = 6$ ($\Delta t = 1$), where the pole plus continuum result misses the data by the factor of 1.8. In Table 1 we compare $\chi^2/dof$ obtained for different time ranges for the two models.

We see that the pole plus continuum model fails when the shortest time is tested. It fails in the region where it is expected to perform the best, since it originates from the asymptotic high energy form of spectral density. One might argue that the pole plus continuum model fails because of the large lattice artifacts, which are the worst in the short time region. However, the same lattice artifacts should plague the three pole model just as badly. The fact, that they do not, can hardly be explained away as a mere coincidence.

At this point we would like to stop and ask some very relevant questions. Are we trying to do the impossible? Given that there is some finite statistical noise, is it possible to distinguish the three pole form of the spectral density from the pole plus continuum? Does our algorithm work for more than three poles? A very careful analysis has to be performed to come to any conclusions. Our goal is twofold: to test our minimization algorithm and to look for more differences between the three pole and pole plus continuum

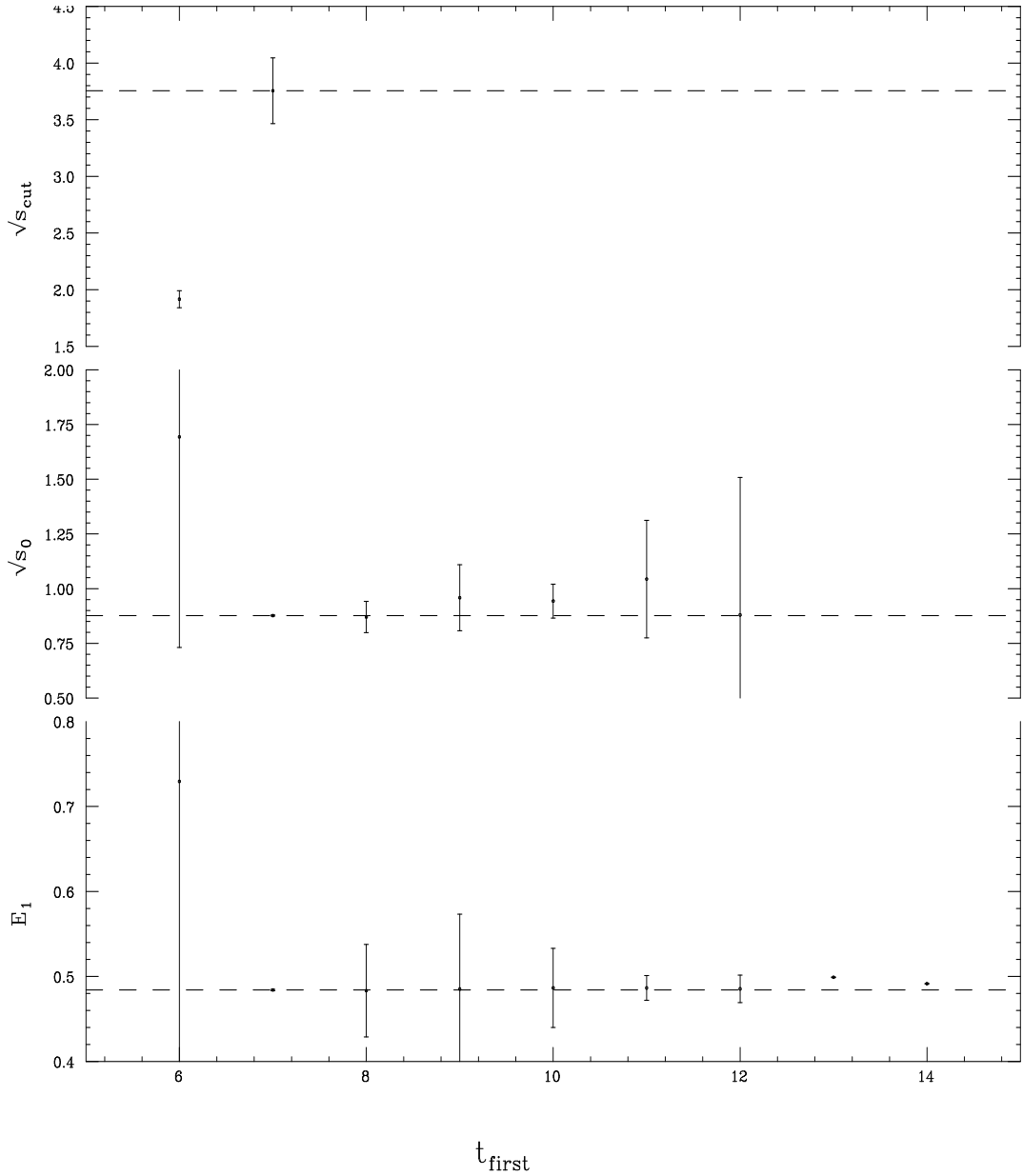

Figure 5: The energy $E_1$ of the ground state, the continuum threshold $\sqrt{s_0}$ and cut-off $\sqrt{s_{cut}}$ as a function of the first time slice $t_{first}$ used in the fit. $t_{last} = 20$ for all the fits. The dashed line shows the result for $t_{first} = 7$. The longest range fit with $t_{first} = 6$ does not give satisfactory results for the pole plus continuum model. The hopping parameter $\kappa = 0.146$.

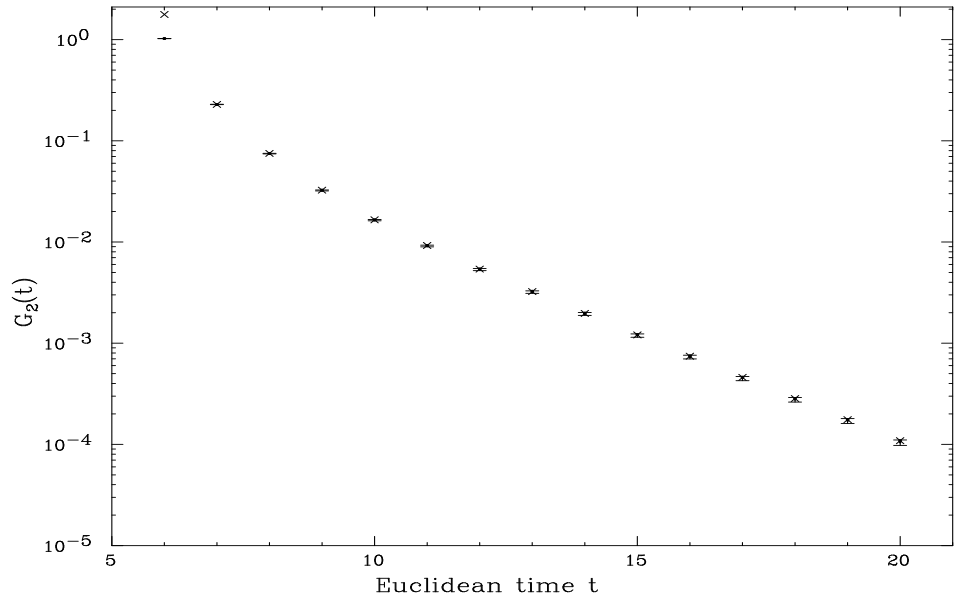

Figure 6: Two-point correlation function $G_2(t)$ (error bars) is fit by the three pole model $G_p(t)$ ('o') and the pole plus continuum model $G_c(t)$ ('x'). The hopping parameter $\kappa = 0.146$.

models.

We generate some simulated many pole $\tilde{G}_c(t)$ and pole plus continuum $\tilde{G}_p(t)$ data. We choose a set of parameters for several poles or for pole plus continuum and calculate $G_c(t)$ and $G_p(t)$ with the formulae (7) and (6). Then we add some statistical noise $\sigma(t)$. In [6] the question of the time dependence of $\sigma(t)$ was discussed. The conclusion was that, while we do not know how $\sigma(t)$ behaves in the short time region, it is reasonable to expect exponential decay given by the pion mass in the long time region. We choose $\sigma(t)$ to have two components. The first one is proportional to the data, the second one is proportional to the ground state signal:

$$\sigma_p(t) = \alpha \tilde{G}_p(t) + \beta e^{-E_1 t}. \tag{17}$$

The coefficients $\alpha$ and $\beta$ are chosen to make the covariance matrix for the generated data to be as close as possible to the covariance matrix of the lattice data.

We test our minimization algorithm on a number of sets of generated data. It is shown to work well to fit the data with up to five poles and the pole plus continuum data for the parameters similar to the real lattice data. The parameters obtained by minimizing $\chi^2$ are same, within the error bars, as the "seed" parameters used to generate the data.

Now we can address the second issue and further our comparison of the three pole and pole plus continuum models. We will try to mimic the lattice data with the data generated by the two models and see which one does a better job. We use the parameters obtained from fitting the lattice data $G_2(t)$ to generate $G_c(t)$ and $G_p(t)$. The only difference is that the parameters for the continuum are taken from $t_{first} = 7$ ($\Delta t = 2$) fit, since at $t_{first} = 6$ ($\Delta t = 1$) the continuum model does not fit the lattice data. We fit the three sets of data with the two spectral density functions. The results for the two longest time ranges with $t_{first} = 6, 7$ ($\Delta t = 1, 2$) are given in Table 4. Already some conclusions can be made. We see that the genuine three pole data can be fit with the pole plus continuum model and visa versa provided we do not test high energy region and do not include the $t = 6$ ($\Delta t = 1$) time slice.

Another interesting result is shown in Table 5. We fit the three sets of data with four pole model. It leads to some real improvement for the pole plus continuum simulated data. In case of lattice data and three pole simulated data the results can be interpreted as the absence of the fourth pole.

| data | $\Delta t$ | $E_1$ | $E_2$ | $E_3$ | $\chi^2/dof$ | $E_1$ | $\sqrt{s_0}$ | $\sqrt{s_{cut}}$ | $\chi^2/dof$ |
|---|---|---|---|---|---|---|---|---|---|
| lattice | 1 | 0.486 | 1.16 | 2.14 | 0.53 | 0.73 | 1.73 | 1.96 | 122 |
|  | 2 | 0.485 | 1.13 | 2.09 | 0.59 | 0.484 | 0.88 | 3.76 | 0.55 |
| 3 p. | 1 | 0.48 | 1.14 | 2.18 | 1.10 | 1.11 | 1.85 | 1.96 | 26.7 |
|  | 2 | 0.48 | 1.10 | 2.01 | 1.17 | 0.48 | 0.86 | 3.64 | 1.13 |
| p.c. | 1 | 0.52 | 1.32 | 3.68 | 33.2 | 0.48 | 0.87 | 3.76 | 0.90 |
|  | 2 | 0.48 | 1.10 | 2.00 | 1.18 | 0.48 | 0.87 | 3.73 | 1.16 |

Table 4: Three sets of data are fit with two different forms of the spectral density. The left side of each table contains three energies of the three pole fit, the right side contains parameters of the pole plus continuum fit. The first table is for the lattice data (lattice), the second one is the data generated by the three pole model (3 p.), and the third one is the data generated by the pole plus continuum model (p.c.). $\Delta t$ is the distance from the source in the time direction. $\Delta E_i = E_i - E_1$.

| data | $\Delta t$ | $E_1$ | $E_2$ | $E_3$ | $E_4$ | $c_1^2$ | $c_2^2 e^{-\Delta E_2 \Delta t}$ | $c_3^2 e^{-\Delta E_3 \Delta t}$ | $c_4^2 e^{-\Delta E_4 \Delta t}$ | $\chi^2$ |
|---|---|---|---|---|---|---|---|---|---|---|
| lattice | 1 | 0.49 | 1.15 | 2.13 | 13.9 | 0.15 | 0.57 | 0.94 | 0.006 | 4.69 |
|  | 1 | 0.49 | 1.16 | 2.14 | - | 0.15 | 0.52 | 1.00 | - | 4.75 |
| 3 p. | 1 | 0.48 | 1.10 | 2.02 | 7.98 | 0.14 | 0.45 | 0.98 | 0.07 | 9.34 |
|  | 1 | 0.48 | 1.14 | 2.18 | - | 0.15 | 0.53 | 0.98 | - | 9.88 |
| p.c. | 1 | 0.48 | 1.08 | 1.90 | 5.46 | 0.14 | 0.39 | 0.95 | 1.38 | 8.46 |
|  | 1 | 0.52 | 1.32 | 3.68 | - | 0.14 | 0.88 | 1.98 | - | 297 |

Table 5: The three sets of data are fit with four and three poles spectral density. The first table is for the lattice data (lattice), the second one is the three pole model (3 p.), and the third one is the pole plus continuum model (p.c.). $\Delta t$ is the distance from the source in the time direction. $\Delta E_i = E_i - E_1$.

| data | $\Delta t$ | $E_1$ | $E_2$ | $E_3$ | $c_1^2$ | $c_2^2 e^{-\Delta E_2 \Delta t}$ | $c_3^2 e^{-\Delta E_3 \Delta t}$ | $\chi^2$ |
|---|---|---|---|---|---|---|---|---|
| lattice | 4 | 0.48 | 1.27 | 3.61 | 0.15 | 0.07 | 4E-06 | 6.18 |
|  | 4 | 0.48 | 1.27 | - | 0.15 | 0.07 | - | 6.17 |
| 3 p. | 4 | 0.48 | 1.17 | 1.92 | 0.14 | 0.08 | 2E-05 | 7.45 |
|  | 4 | 0.48 | 1.17 | - | 0.14 | 0.08 | - | 7.45 |
| p.c. | 4 | 0.48 | 1.06 | 1.70 | 0.14 | 0.06 | 0.02 | 7.47 |
|  | 4 | 0.48 | 1.19 | - | 0.14 | 0.08 | - | 105 |
|  | 5 | 0.48 | 1.07 | 1.80 | 0.14 | 0.03 | 0.005 | 7.57 |
|  | 5 | 0.48 | 1.14 | - | 0.14 | 0.04 | - | 21 |

Table 6: The three sets of data are fit with three and two poles spectral density. The first table is for the lattice data (lattice), the second one is the three pole model (3 p.), and the third one is the pole plus continuum model (p.c.). $\Delta t$ is the distance from the source in the time direction. $\Delta E_i = E_i - E_1$.

Let us see what else can be learned with these simulations. We can make the following argument. The lattice data and both simulated data are fit by the continuum and three pole models when $t_{first} = 7$ ($\Delta t = 2$) and the fit parameters of these three fits are the same within the error bars. Suppose the assumption, we ruled out before, that lattice artifacts are responsible for the failure of the pole plus continuum to fit the data in the short time range, is true. Suppose the lattice data does correspond to the pole plus continuum spectral density, which just happened to be fit equally well by the three pole model. Now we fit these three sets of data with three and two pole models for several time ranges with increasing $t_{first}$. Based on the above assumption one would expect the results of these fits to be similar as well. The fit parameters for the three sets of data are given in the Table 6 for $t_{first} = 9, 10$ ($\Delta t = 4, 5$). The lattice and three pole simulated data show similar behaviour: the third pole drops out and the data is fit with two poles for $t_{first} = 9$ ($\Delta t = 4$). In the stark contrast the third pole remains important for the fit of the pole plus continuum simulated data even for $t_{first} = 10$ ($\Delta t = 5$).

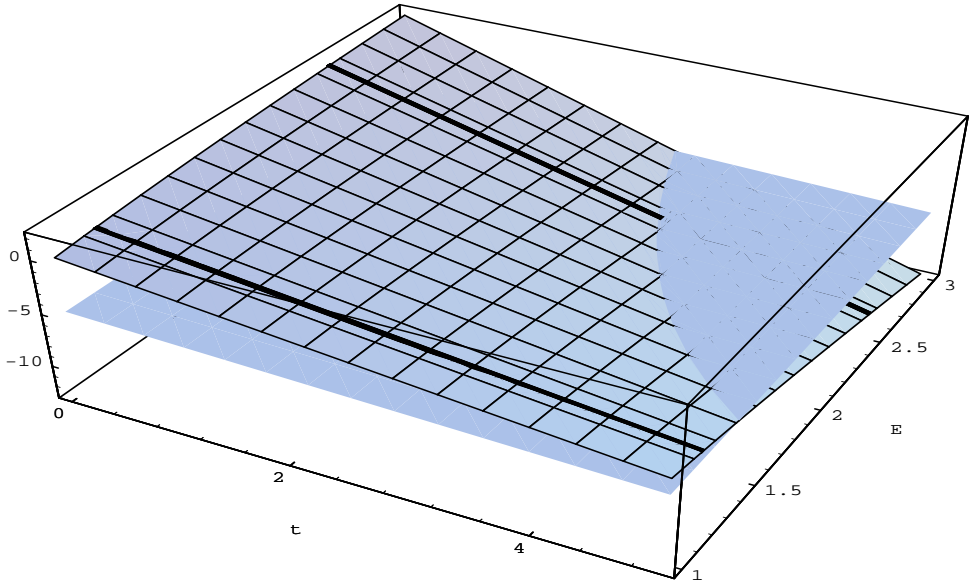

Figure 7: We plot $\log\left(f(E)e^{-Et}\right)$ as a function of $E$ and $t$. The surface with mesh lines to $f(E) = E^2$, and the two rays correspond to $f(E) = c_1^2 \delta(E - E_1) + c_2^2 \delta(E - E_2)$. The surface without mesh lines represents the statistical error.

This difference can be explained as follows. Since for the continuum simulated data there is in reality no third state, the fitting parameters of this state have greater flexibility and adjust as the portion of the spectrum remaining above the statistical error decreases. Fig. 7 is designed to make this point more transparent. Suppose we fit a many pole data. In that case, if a contribution of a state goes beyond the statistical error, it is gone and there is no trace of it. For the continuous spectral density the high energy contribution disappears gradually. If this data is fit with the many pole model the energy of the states

| data | $\Delta t$ | $c_1^2$ | $E_1$ | $c_{cont}$ | $\sqrt{s_0}$ | $\chi^2/dof$ |
|---|---|---|---|---|---|---|
| | 2 | 0.15 | 0.484 | 1.25 | 0.87 | 0.55 |
| lattice | 3 | 0.14 | 0.483 | 1.24 | 0.87 | 0.61 |
| | 4 | 0.15 | 0.485 | 1.56 | 0.96 | 0.58 |
| | 2 | 0.14 | 0.48 | 1.3 | 0.86 | 1.13 |
| 3 p. | 3 | 0.14 | 0.48 | 1.3 | 0.86 | 1.27 |
| | 4 | 0.14 | 0.48 | 4.1 | 1.08 | 1.07 |
| | 2 | 0.14 | 0.48 | 1.3 | 0.87 | 1.16 |
| p.c. | 3 | 0.14 | 0.48 | 1.3 | 0.86 | 1.39 |
| | 4 | 0.14 | 0.48 | 1.3 | 0.89 | 1.15 |

Table 7: The three sets of data are fit with the pole plus continuum spectral density. The first table is for the lattice data (lattice), the second one is the three pole model (3 p.), and the third one is the pole plus continuum model (p.c.). $\Delta t$ is the distance from the source in the time direction.

can be shifted to adjust to the data.

A similar difference arises when we fit the three sets of data with the pole plus continuum spectral density (Table 7). The three sets of data are fit with the same parameters for $t_{first} = 7$ and 8 ($\Delta t = 2$ and 3). For $t_{first} = 9$ ($\Delta t = 4$) the fit parameters for the lattice and three pole data suddenly change, whereas the pole plus continuum is fit with the same parameters as before.

## 5  Spatial Correlation Functions

Spatial correlation functions $S_2(x)$ have the following spectral representation:

$$S_2(x) = \frac{1}{(4\pi)^2 x} \int ds f(s) K_1(\sqrt{s}x). \tag{18}$$

Since we work in the Euclidean space all four coordinates are equivalent. But the influence of the hard wall boundary condition used for quarks in the time direction is not as easy to account for as the periodic boundary condition in the spatial directions. Therefore we place our source on the central time slice and measure $S_2(x)$ only for spatial separations.

We fit the spatial correlation functions measured on the lattice for $\pi$ and $\rho$ and for $\kappa = 0.146, 0.149$ with the many pole (7) and pole plus continuum (6) forms of the spectral density. With these forms of the spectral density equation (18) yields the many pole model fit function

$$S_p(x) = \frac{1}{4\pi^2 x} \sum_n \lambda_n^2 E_n K_1(E_n x) \tag{19}$$

and the pole plus continuum model fit function

$$S_c(x) = \frac{1}{4\pi^2 x} \lambda_1^2 E_1 K_1(E_1 x) + \frac{c_{cont}}{2\pi^2 x^6} \int_{\sqrt{s_0}x}^{\infty} d\alpha \, \alpha^4 K_1(\alpha), \tag{20}$$

where $K_1(\alpha)$ is the modified Bessel function.

So far this section nicely parallels the previous one. However, at this point we have to discuss the two issues that complicate the matter. They are: the lattice anisotropy and the contributions from the images. The effect of anisotropy becomes clear if we consider the hopping parameter expansion for the noninteracting quarks. Various equidistant points can be reached by different numbers of steps (links) on the lattice. This effect disappears at large separations, but it is of little consolation if our intention is to examine the short distance behaviour of the propagators.

The images are the consequence of the periodic boundary conditions for quarks in the spatial directions (Fig 8). In fact, the function $S_2(x)$ we measure on the lattice is a sum of infinite number of terms:

$$S_2(x) = \sum_{\vec{n}} S_2^{\infty}(x + \vec{n}L). \tag{21}$$

Each term is the infinite volume correlator $S_2^{\infty}(x + \vec{n}L)$ connecting the source to all sinks obtained by shifting the primary sink in any of the spatial directions by a multiple of the lattice size $L$ (Fig. 8). In practice, we only need to consider 8 sinks: the primary one and its 7 closest images.

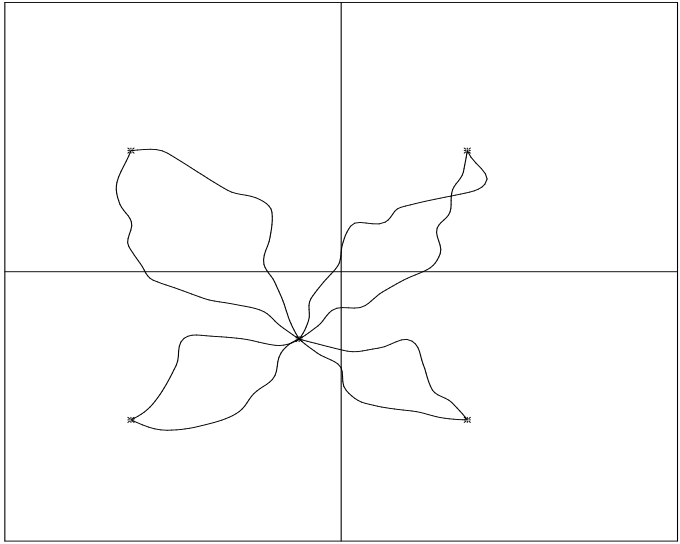

Figure 8: The propagators from the source to the sink and its three nearest images are shown schematically.

Note that one does not encounter this problem with the zero-momentum correlation function. The summation of the lattice correlation function $S_2(x)$ over the points of one cell is equivalent to the summation of the infinite volume correlation function $S_2^\infty(x)$ over infinite space. ($G_2(t) = \sum_{\vec{x}} S(\vec{x}, t)$, if we include time and write $x$ explicitly as $(\vec{x}, t)$.)

In papers [17, 9] the authors proposed a cure for both problems. The image correction has to be done by numerically subtracting the image contributions. It is supposed to be done iteratively by approximately correcting for images using an appropriately defined parametric curve, least squares fitting the parameters to the corrected data, and iterating to self-consistency. This procedure yields a smooth universal curve at large distances. The authors in [9] then find the correlators for several quark masses, extrapolate to the physical limit, and fit the curve again with $S_c(x)$ corresponding to the pole plus continuum form of the spectral density.

To account for the lattice anisotropy they normalize $S_2(x)$, corrected for the images, by the meson correlator in the case of free massless quarks. This was shown to reduce the anisotropy significantly. Further reduction of the effect of the lattice anisotropy proposed in [9] is based on the following observation. Using the free quark lattice correlator, which can be computed analytically for infinite space (thereby eliminating the images), they noticed that anisotropic effects were relatively small for the points lying in the diagonal direction from the source and within a cone surrounding the diagonal with opening angle $\theta \leq \arccos{(0.9)} \approx 26°$.

The difference between our case and theirs is the physical size of the lattice. In our case it is $\approx 1.2$ fm, in their case it is $\approx 2.7$ fm. As a result, we can not perform simple image correction based on the ground state contribution. We would have to do the full-fledged fitting in order to correct for the images.

In this paper we account for the images explicitly. That is, we fit the lattice data $S_2(x)$ with the function $\bar{S}_p(x)$ ($\bar{S}_c(x)$) equal

$$\bar{S}_p(x) = \sum_{\vec{n}} S_p(x + \vec{n}L), \tag{22}$$

where the summation is over eight closest to the source points, corresponding to $\vec{n} = $ (0,0,0), (1,0,0), (0,1,0), (0,0,1), (1,1,0), (1,0,1), (0,1,1), (1,1,1).

Since we account for the images explicitly, we can not normalize $S_2(x)$ by the free massless quark correlators. Therefore, we use even more stringent requirement $\theta \leq \arccos{(0.95)} \approx 18°$ in the attempt to eliminate anisotropic effects. We end up with 17 points that satisfies this criterion. Their distances to the source range from 1.73 to 8.66 in the lattice units. In Fig. 9 we show $S_2(x)$. The effects of the lattice anisotropy and image contributions are apparent. $S_2(x)$ for equidistant points differ by an order of magnitude. But the points within the cone $\theta \leq \arccos{(0.95)}$ form a reasonably smooth curve for short distances. For long distances the non-smoothness is the result of image contributions.

We generated 120 configurations, the number big enough for us to perform the correlated $\chi^2$ fit.

In Fig. 10 the energy of the ground and the first excited states are shown as a function of the first point used in the fit. The results for other hopping parameters and for $\rho$ are not satisfactory. The energy of the first excited state is not determined. The energy of the ground state is stable in the long distance tail, but deviates significantly from the value obtained in the previous section. Unfortunately, the spatial correlation functions $S_2(x)$ can not be fit nearly as well as the zero momentum correlation functions $G_2(t)$. The values of $\chi^2/dof$ are outrageously high. They are given in Table 8 for some of the fitting ranges.

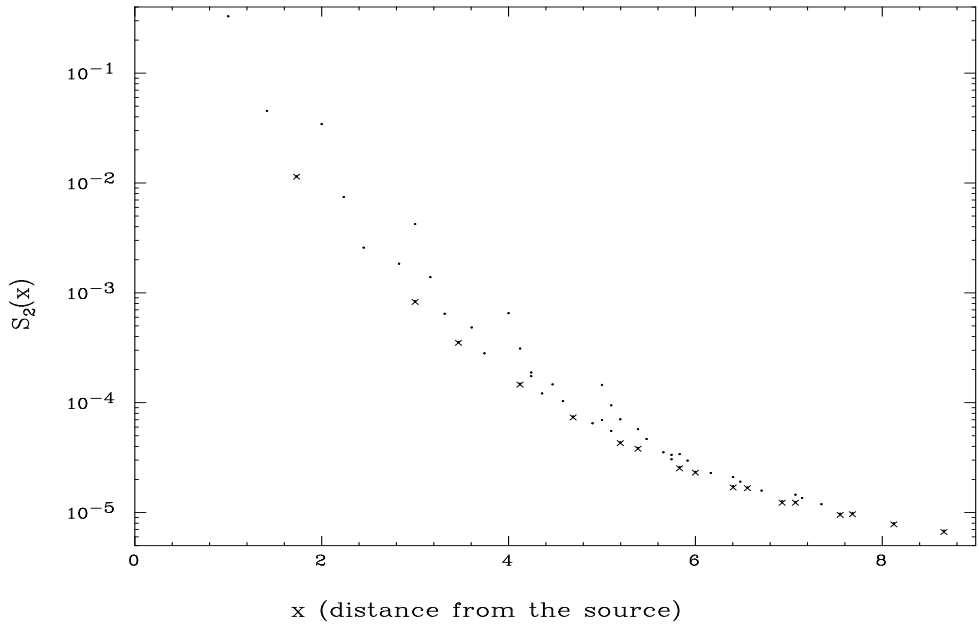

Figure 9: The two-point spatial correlation function $S_2(x)$ in the pseudoscalar channel for $\kappa = 0.146$. The points located within the cone $\theta \leq \arccos(0.9) \approx 26°$ are marked with 'x'.

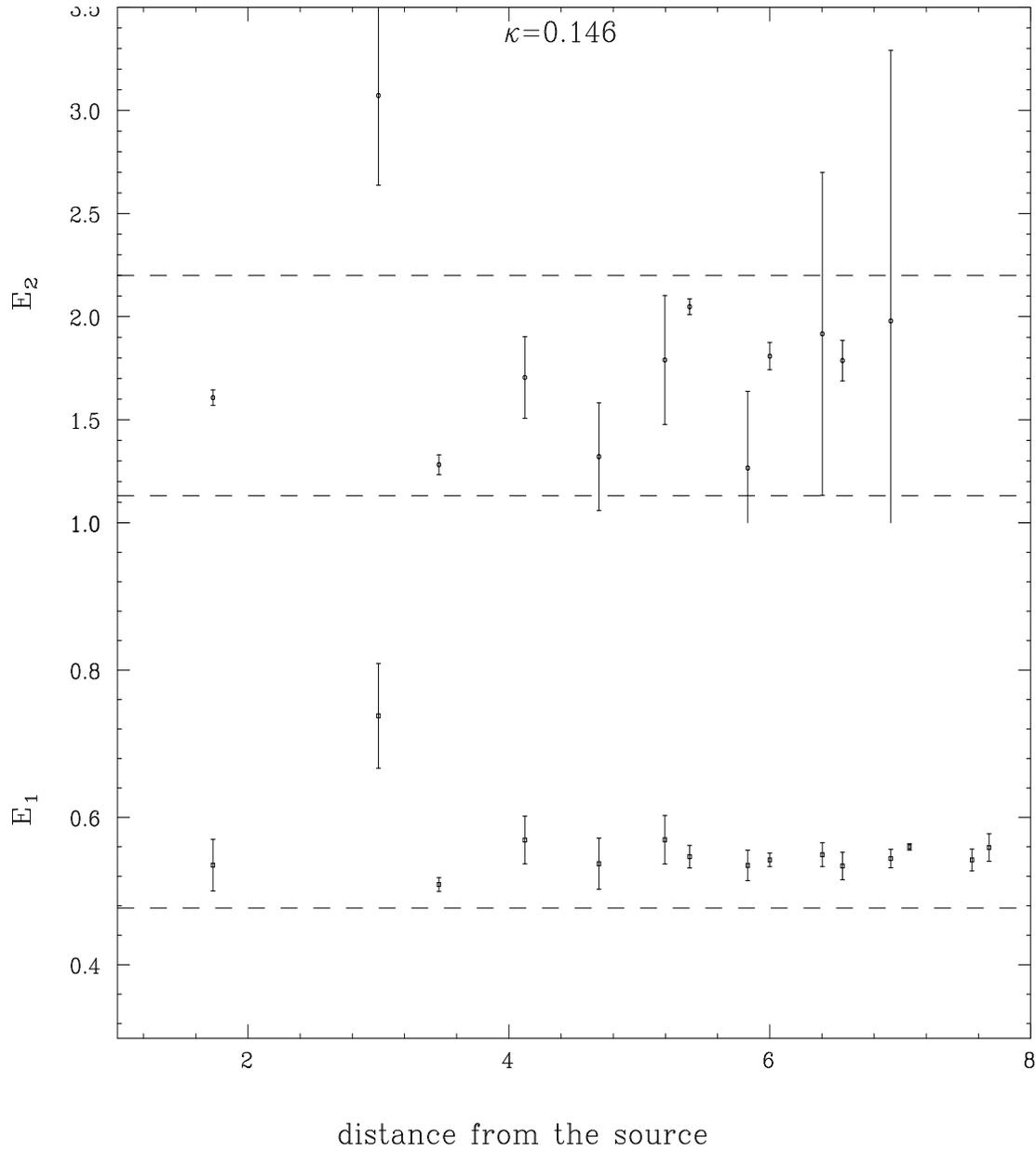

Figure 10: The energies of the two state fit of the spatial correlation function $S_2(x)$ as a function of the distance from the source to the first point included in the fit. The dashed lines show the values of the energies of the three states obtained for the zero momentum two-point correlation function $G_2(t)$. $\pi$ channel; hopping parameter $\kappa = 0.146$.

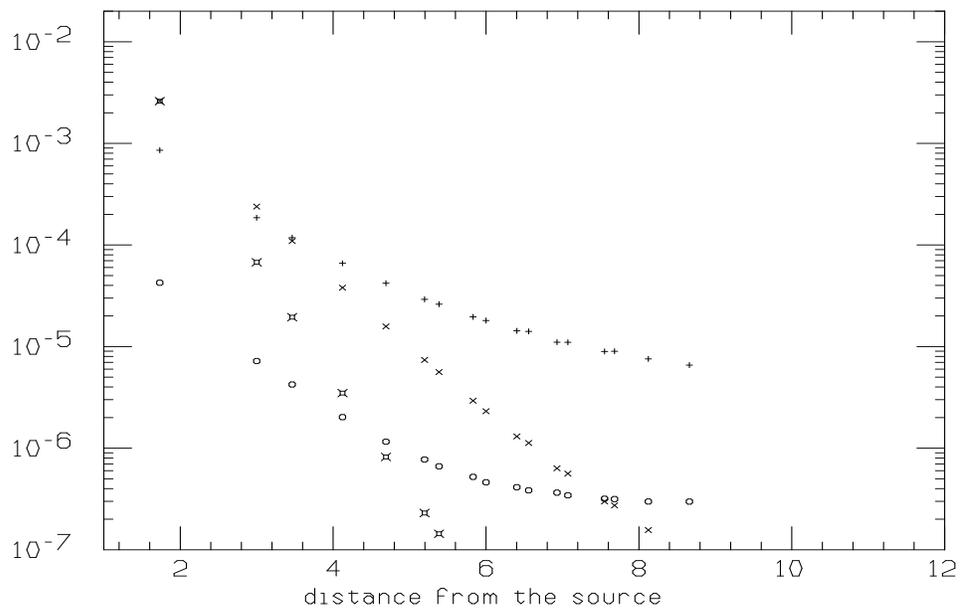

Figure 11: The contributions of the three states (vertical, diagonal, and fancy crosses) to the two-point spatial correlation function $S_2(x)$ to be compared with the statistical error ('o') of $S_2(x)$.

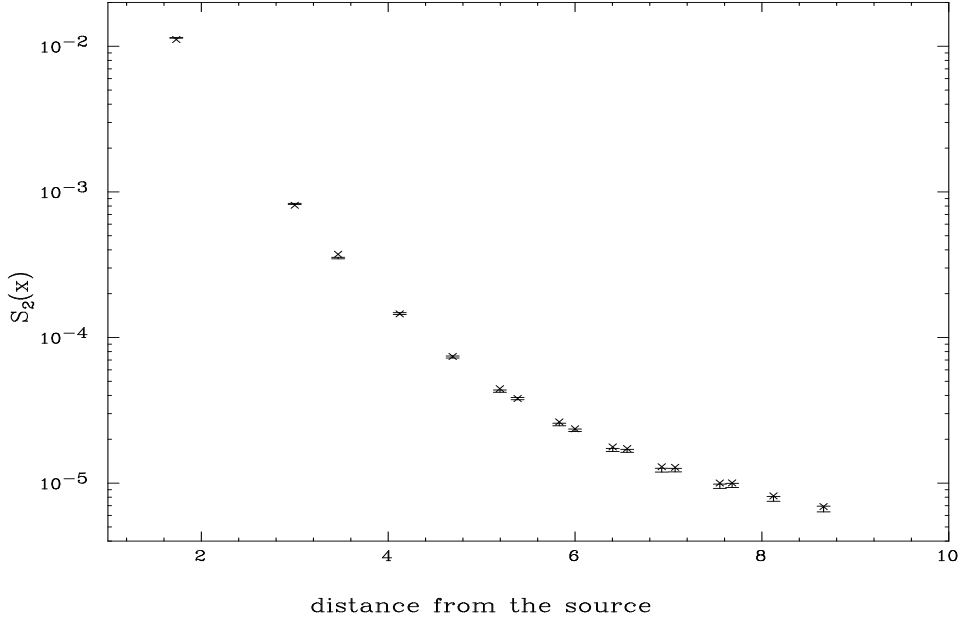

Figure 12: The two-point spatial correlation function $S_2(x)$ (error bars) fit with $S_p(x)$ for two poles ('x'). $\pi$ channel; hopping parameter $\kappa = 0.146$.

The contribution of the third excited state can not be detected. If the three pole form is used to fit the data the third state comes out exactly the same as the second one. To see if the statistical error is the cause, in Fig 11 we plot the contribution to $S_p(x)$ of the three states computed with the parameters $\lambda_n$ and $E_n$ obtained for $G_2(t)$. There are four points for which the contribution of the would be third state is above the statistical error. Based on the analysis in the previous section the four points should be enough to extract a contribution of a state.

In Fig. 12 the spatial two point correlation function $S(x)$ for $\pi$ and $\kappa = 0.146$ is shown fit with $S_p(x)$ corresponding to the two pole spectral density. The high values of $\chi^2/dof$ (Table 8) are the result of the very small statistical errors on the data.

| $n_{first}$ | 1 | 2 | 3 | 5 | 7 | 9 | 11 | 13 | 15 |
|---|---|---|---|---|---|---|---|---|---|
| $\chi^2/dof$ | 80 | 47 | 5.6 | 2.0 | 1.6 | 2.1 | 1.8 | 2.9 | 1.0 |
| # poles | 2 | 2 | 2 | 2 | 2 | 2 | 2 | 1 | 1 |

Table 8: $\chi^2$ per degree of freedom for the fits with the first point $n_{first}$ included in the fit. The number of poles is also shown.

The attempt to fit $S(x)$ with the pole plus continuum model fails completely. What is supposed to be the continuum threshold $s_0 > E_1$ came out very close to 0.

Apparently, the systematic error caused by the lattice anisotropy exceeds the statistical error. The anisotropy does not affect as much the zero momentum correlation function since the summation over space smooths it out.

# 6  Where Do We Go From Here

It is of interest to investigate a question which is the better way to examine the excited states, or in other words to go up in energy in spectral density. There can be two approaches. One is to increase the statistics to lower the statistical error and to dig out the contribution from higher energy region from under the statistical error. Another one is to decrease the lattice spacing with the corresponding increase in the lattice size to keep the physical size constant. In principle decreasing the lattice spacing only in the time direction causes some problems as far as recovering the continuum limit result [18]. But if we are only interested in the form of the spectral function on the lattice that should not concern us. The question is which one is cheaper in terms of computer time. We explained above that to detect a state we need at least three time slices with the signal from this state above the statistical error. Consider a simple example. The next excited state $|n>$ goes below the statistical level at the first time slice (see Fig. 13). In the first scenario we would have to decrease the statistical error so that this signal is above it for at least three time slices. We need to increase the number of the configurations from $N_1$ to $N_2$. We assume that the statistical error $\Delta G_2(t)$ is inversely proportional to the number of configurations $N$:

$$\Delta G_2(t) = \frac{1}{\sqrt{N}} er(t), \tag{23}$$

where $er(t)$ is some in general unknown function of time $t$. The above condition can now be written as follows:

$$\frac{1}{\sqrt{N_1}} er(1) = c_n e^{-E_n} \tag{24}$$
$$\frac{1}{\sqrt{N_2}} er(4) = c_n e^{-4E_n}$$

Which gives us the following condition for the increase in the number of configurations:

$$N_2/N_1 = e^{6E_n} (er(4)/er(1))^2 \tag{25}$$

We need to know at least something about the time dependence of the statistical error $er(t)$. In [6] it was shown that in the large $t$ limit $er(t) \sim e^{-m_\pi t}$. As far as short $t$ region we do not have any theoretical predictions. We will use our data for $G_2(t)$ for $\pi$ and $\kappa = 0.146$ to get a rough estimate of $N_2/N_1$. The statistical error in this channel in the short time region is fit well as

$$er(t) \sim e^{-1.8t}. \tag{26}$$

The energy of the first and second excited states determined in Section 4 are 1.2 and 2.2, correspondingly. We will estimate the energy of the fourth excited state as $2.4 < E_4 < 2.8$. Within this energy range the number in question $N_2/N_1 \sim 40 \div 400$.

In the second scenario we assume that the statistical error will be the same for different lattice spacing and the same number of configurations. Then to get additional three time slices with the signal of the excited state in question we will merely have to quadrupole the temporal extension of the lattice. The time required to generate the gauge configurations will increase by exactly this factor. To calculate the quark propagators inversions of the sparse matrix performed. The inversion time for a sparse matrix can also be expected to grow proportionally to the lattice size. This dependence has been observed empirically in our calculations.

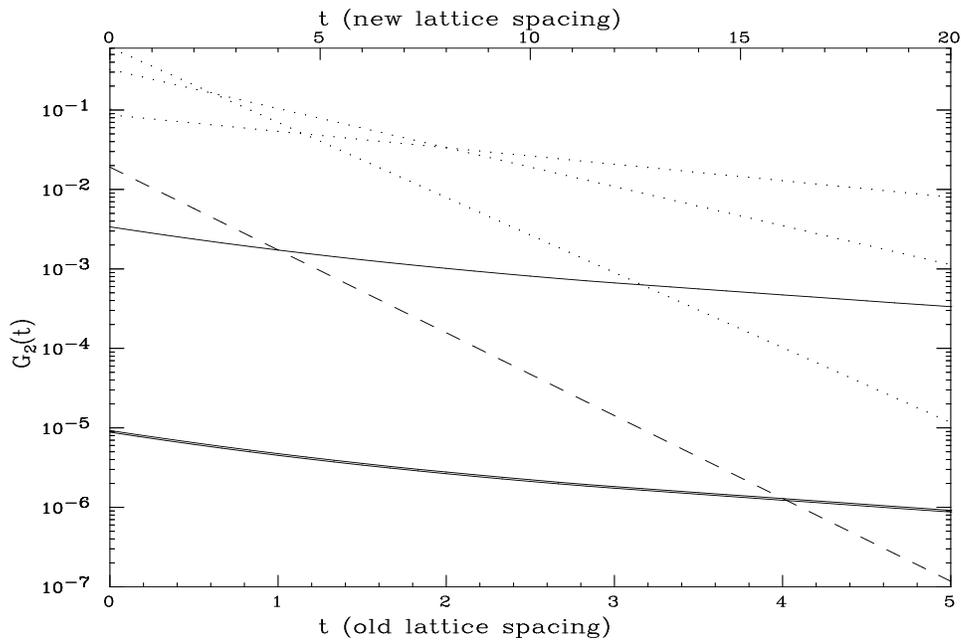

Figure 13: Individual contributions of several states to the two-point correlation function to be compared to the statistical error of the two-point correlation function $G_2(t)$. The contribution of the state under consideration $n$ is shown with dashed line, all others are shown with dotted lines. The statistical error for $N_1$ and $N_2$ configurations are shown, correspondingly, with the single and double solid lines.

Therefore, we have a factor of 4 vs. a factor of $40 \div 400$. We see that decreasing the lattice spacing in the time direction is much more promising way of looking at the high energy part of the spectral density on the lattice. Note that should we try to keep the lattice isotropic and increase it in all four dimensions it would cost us a factor of 256 which is about the same as the cost of the reduction of the statistical error. This should not be surprising.

# 7 Conclusion

In this paper we examine the spectral density function on the lattice. We use spatial and definite momentum two-point correlation functions measured on the lattice in the pseudoscalar and vector channels for three values of the quark mass. We show that with the data of reasonable accuracy ( order of 100 configurations) one can extract the spectral density function using a correlated fit of zero-momentum two-point correlation functions. This paper supports the conclusion reached in the previous paper [11] that within the accessible energy range the spectrum in the channels under consideration consists of three poles. Spatial two-point correlation functions are shown not to be well suited for this purpose due to the strong anisotropic effects.

Further investigation with smaller lattice spacing is needed to have a better resolution of already avaliable portion of the spectrum as well as an ability to examine higher energies.